# Early Transient Period in the Evolution of the Area of a Passive Front Propagating in a Strong Turbulence

V.A. Sabelnikov[1,2] and A.N. Lipatnikov[3]

[1]ONERA – The French Aerospace Laboratory, F-91761 Palaiseau, France,
[2]Central Aerohydrodynamic Institute (TsAGI), 140180 Zhukovsky, Moscow Region, Russian Federation,
[3]Department of Applied Mechanics, Chalmers University of Technology, 412 96, Gothenburg, Sweden

## ABSTRACT

Influence of statistically stationary, homogeneous, and isotropic turbulence on the mean area of a passive self-propagating front and, hence, on the rate of fluid consumption by the front is analysed in the case of asymptotically high turbulent Reynolds number $\mathrm{Re}_L$ and asymptotically high ratio of the Kolmogorov velocity to a constant speed $u_0$ of the front. By considering an early stage of the front evolution, the mean (over the studied early stage) front area and consumption velocity $\overline{u}_T$ are analytically determined. The analysis shows that the mean $\overline{u}_T$ is proportional to the rms turbulent velocity, which characterizes large-scale turbulent eddies, even if the instantaneous rate of an increase in the front area is mainly controlled by the smallest Kolmogorov eddies. A straightforward dependence of the mean area or $\overline{u}_T$ on the Kolmogorov velocity, length, and time scales vanishes due to the following physical mechanism. At high $\mathrm{Re}_L$, turbulent stretching created by small-scale eddies increases the front area exponentially with time, whereas a volume bounded by the leading and trailing edges of the front grows significantly slower. Therefore, after a short time, the volume is tightly filled by the front and the mean distance between opposed front elements becomes small with respect to Kolmogorov length scale. Subsequently, such front elements collide, thus, reducing the front area and limiting the mean $\overline{u}_T$.

**Introduction**

The problem of passive front propagation in randomly advected media (Chertkov and Yakhot, 1998; Kerstein and Ashurst, 1992; Mayo and Kerstin, 2008) is straightforwardly relevant, e.g., to autocatalytic reactions in liquids (Aris, 1999; Shy et al., 1992), and, as a toy problem, to various phenomena, ranging from turbulent combustion (Chaudhuri et al., 2012) and deflagration-to-detonation transition (Poludnenko et al., 2011) under terrestrial conditions to evolution of thermonuclear Ia supernovae (Gamezo et al., 2003 and 2004) in the Universe. Historically, the problem attracted much attention since 1940s when significant acceleration of flame propagation by turbulence was found. The phenomenon was explained by Damköhler (1940) and Shelkin (1943) who highlighted random advection of a flame by turbulent flow and reduced the influence of the turbulence on the flame to an increase in the area of the flame surface wrinkled due to large-scale velocity fluctuations. Following those pioneering ideas, various models of front propagation in a turbulent flow express the mean turbulent consumption velocity $\bar{u}_T$ (i.e., the mean mass rate of consumption of a fluid per unit area of the surface of the mean front, normalized using the fluid density upstream of the front) to be a function of the front speed $u_0$ and the rms turbulent velocity $u'$, with a ratio of $\bar{u}_T/u_0$ being controlled by the mean increase in the front area. For instance, such expressions are widely used in the turbulent combustion literature (Poinsot and Veynante, 2005). Moreover, recent Direct Numerical Simulation (DNS) study (Yu et al., 2015) of self-propagation of a passive interface[1] in constant-density turbulence showed a linear relation between $\bar{u}_T$, $u_0$, and $u'$ at least at $0.5 \leq u'/u_0 \leq 10$.

However, in spite of long-term investigations of propagation of a front (e.g., a flame) in randomly advected media (e.g., turbulence), physical mechanisms that result in the aforementioned linear relation do not seem to be fully clarified. To show the issue let us first assume that (i) the influence of turbulence on a front is reduced to an increase in its surface area due to wrinkles caused by turbulent eddies and (ii) the classical Kolmogorov theory (1941) is applied to characterize the turbulence and its influence on the surface area. Under such assumptions, the surface area, at least in the case of a material surface, is expected to be mainly created due to local stretch rates produced by the small-scale eddies (Batchelor, 1952) whose statistical properties are controlled by the mean dissipation rate $\bar{\varepsilon} = 2\nu \overline{S_{ij} S_{ij}}$ and kinematic viscosity $\nu$. Here, $S_{ij} = 0.5(\partial u_i / \partial x_j + \partial u_j / \partial x_i)$ is the rate-of-strain tensor, $u_i$ is the $i$-th component of velocity vector $\mathbf{u}$, and summation convention applies to repeated indexes $i$ and $j$. Accordingly, an increase in the mean area $\bar{A}_F$ of the front by turbulent eddies and, hence, a ratio of $\bar{u}_T / u_0$ is expected to be mainly controlled by a single turbulence characteristic such as $\bar{\varepsilon}$, rather than $u'$.

---

[1] The consumption velocity depends not only on $u'$ and $u_0$, but also on $L$ and the laminar wave thickness in the case of a reaction wave of a finite thickness (Yu and Lipatnikov, 2017a and 2017b) and, in particular, in premixed flames (Lipatnikov and Chomiak, 2002). However, such effects are beyond the scope of the present communication, which is restricted to an infinitely thin front.

Then, in the case of statistically stationary front propagation, $\bar{u}_T$ may depend on $u_0$, $\bar{\varepsilon}$, and $\nu$, i.e., for dimensional reasoning, $\bar{u}_T / u_0 = f(\nu \bar{\varepsilon} / u_0^4)$, with the dissipation rate and viscosity jointly controlling the scales of the smallest eddies associated with the dominant contribution to the front area increase. In such a case, however, a dependence of $\bar{u}_T$ on $u'$ should be accompanied by the qualitatively opposite dependence of $\bar{u}_T$ on an integral length scale $L$ of the turbulence, because $\bar{\varepsilon} \propto u'^3 / L$ within the framework of the Kolmogorov (1941) theory. Accordingly, a scaling like $\bar{u}_T \propto u'$ at $u_0 \ll u'$, which is widely accepted and was confirmed in the aforementioned DNS study (Yu et al., 2015), cannot be obtained by reducing the influence of turbulence on the front to its stretching by the small-scale velocity fluctuations. Influence of the large-scale velocity fluctuations on the front motion should also be taken into account in order to arrive at $\bar{u}_T \propto u'$ at $u_0 \ll u'$.

To resolve the problem, turbulent entrainment, which is controlled by large-scale eddies is commonly highlighted, with small-scale characteristics of any surface (material of self-propagating) being assumed to be adjusted to the influence of large-scale turbulent eddies on the surface (Tsinober, 2009). Subsequently, the fractal concept (Sreenivasan et al., 1989) is invoked to describe the surface characteristics at various scales and in particular to yield $\bar{u}_T \propto u'$ at $u_0 \ll u'$ provided that the fractal dimension $D=7/3$ and the inner cut-off scale $e_{out}$ is equal to the Gibson length scale $L_G$ (Peters, 1986; Niemeyer and Kerstein, 1997). However, the concept does not reveal a physical mechanism that results in adjustment of small-scale characteristics of the surface to its large-scale characteristics. Moreover, experimental data obtained from premixed turbulent flames support neither $D=7/3$ nor $e_{out}=L_G$, as reviewed elsewhere (Lipatnikov, 2012).

Accordingly, the primary goal of the present communication is to hypothesize a specific physical mechanism that reconciles (i) a widely recognized scaling of $\bar{u}_T \propto u'$ at $u_0 \ll u'$, (ii) the concept of turbulent entrainment, and (iii) a well-recognized paradigm that reduces the influence of turbulence on a front to an increase in the front area by turbulent eddies characterized within the framework of the Kolmogorov theory. For this purpose, a simple model problem will be stated and studied analytically in the next section.

Furthermore, because this simple study (i) is restricted to an early stage of front propagation in order to obtain analytical results and (ii) highlights transient effects, another goal of the communication consists in qualitatively discussing a role played by such transient and other effects during a fully-developed stage of front propagation, characterized by statistically stationary consumption velocity and mean wave thickness averaged over a time interval sufficiently long when compared to eddy-turn-over time $\tau_T = L/u'$. These issues are discussed in Sec. III followed by conclusions.

**A simple model problem**

Let us consider an infinitely thin front that propagates at a constant speed $u_0$ in statistically stationary, homogeneous, isotropic turbulence that (i) is not affected by the front, (ii) is characterized by a high turbulent Reynolds number $\mathrm{Re}_L = u'L/\nu \gg 1$ and, therefore, (iii) is described by the Kolmogorov (1941) theory. Moreover, in order to obtain analytical results, let us (i) assume that the Kolmogorov velocity $u_K = (\nu\bar{\varepsilon})^{1/4}$ is much larger than $u_0$ and (ii) address an early stage of the growth of the surface of an initially planar front embedded into the turbulence at $t = 0$. Subsequent evolution of the front will be discussed in the next section.

The following analysis is based on (i) the theory of the growth of a material surface area in the Kolmogorov turbulence, developed by Batchellor (1951), (ii) results of DNS studies (Girimaji and Pope, 1990; Goto and Kida, 2007; Yeung et al., 1990) of this phenomenon, and (iii) the classical theory of turbulent diffusion, developed by Taylor (1935).

On the one hand, if a planar material surface is embedded into the Kolmogorov turbulence normally to the $x$-axis, then, after a short transient time interval $t \geq t_i \approx (2.5 - 3)\tau_K$ during that the surface adapts itself to the turbulent field, the mean (ensemble-averaged) surface area $A_M(t)$ is well known to grow exponentially with time (Batchellor, 1951; Girimaji and Pope, 1990; Goto and Kida, 2007; Yeung et al., 1990), i.e.,

$$A_M(t) = A_0 \exp(\xi t/\tau_K), \tag{1}$$

where $\tau_K = (\nu/\bar{\varepsilon})^{1/2}$ is the Kolmogorov time scale, $\xi$ is constant close to 0.28 (Girimaji and Pope, 1990; Yeung et al., 1990), and $A_0$ is the area of the initial planar material surface at $t = 0$.

On the other hand, the rms dispersion $\Delta_M(t)$ of the $x$-coordinates of points on the surface is well-known to grow linearly with time (Taylor, 1935)

$$\Delta_M(t) \propto u't \tag{2}$$

at $0 < t \ll \tau_T$. It is worth noting that constraints of $t \geq t_i \approx (2.5-3)\tau_K$ and $t \ll \tau_T$ are consistent with one another in the considered case of $\text{Re}_L \gg 1$. It is also worth noting that $\Delta_M(t) \propto \sqrt{u'Lt}$ at $\tau_T \ll t$ (Taylor, 1935), but this limit case is of minor interest for the present analysis, because results obtained in the rest of this section hold at $t \ll \tau_T$.

As argued in (Yeung et al., 1990), Eqs. (1) and (2), which are valid for a material surface, describe also the growth of the area $A_F(t)$ of a passive self-propagating front and the rms front dispersion $\Delta_F(t)$, respectively, provided that (i) $u_K \gg u_0$ and $t \geq t_i \approx (2.5-3)\tau_K$ and (ii) $u' \gg u_0$ and $0 < t \ll \tau_T$, respectively. Thus, at $u_K \gg u_0$ and $t_i < t \ll \tau_T$, the following two equations

$$A_F(t) = A_0 \exp(\xi t/\tau_K), \tag{3}$$

$$\Delta_F(t) \propto u't \tag{4}$$

hold simultaneously. Comparison of Eqs. (1)-(2) with Eqs. (3)-(4) shows that material and self-propagating surfaces that coincide at $t=0$ remain to be close to one another at $0 < t \ll \tau_T$, with the distance between them being smaller than the Kolmogorov length scale $\eta_K = u_K \tau_K$ with a high probability (Yeung et al., 1990). This feature could be attributed to the well-known statistical dominance of positive rates of strain of a material surface in the Kolmogorov turbulence. Because (i) the magnitude of the local velocity normal to a material surface is increased with distance from the surface in the case of a positive local strain rate and (ii) the normal velocity vector $\mathbf{u}_n$ points to the surface, the velocity $|\mathbf{u}_n|$ can be much larger than $u_0 \ll u_K$ at a small distance from the surface, thus, impeding further divergence of the material and self-propagating surfaces.

However, there are fundamental differences between the two surfaces. Indeed, because different elements of a material surface never collide, the area of the material surface grows exponentially and the distance $d$ between different elements of the surface can be very small, as small as we wish. For instance, DNS data by Yeung et al. (1990) show that the distance $d$ is randomly distributed in a wide interval of length scales, see Fig. 6 in the cited paper. On the contrary, elements of a self-propagating surface can collide, thus, consuming the surface area if the local distance between the elements is sufficiently small.

At instant $t$, the mass rate of the consumption of a fluid (e.g. the deficient reactant in the case of a reaction front) by a front is equal to $\rho_d u_0 A_F(t)$, where $\rho_d$ is the fluid density upstream of the front. Accordingly, the fluid volume consumed at the same instant is equal to $u_0 A_F(t)$ and the volume consumed during a time interval $(0,t)$ may be estimated as follows

$$V_F(t) = u_0 \int_0^t A_F(\theta) d\theta \approx u_0 A_0 \int_{t_i}^t \exp\left(\xi \frac{\theta}{\tau_K}\right) d\theta = u_0 \tau_K \xi^{-1} A_0 \left[\exp\left(\xi \frac{t}{\tau_K}\right) - \exp\left(\xi \frac{t_i}{\tau_K}\right)\right]. \quad (5)$$

If $t_i \ll t$, the second term in square brackets may be neglected and we arrive at

$$V_F(t) \approx u_0 \tau_K \xi^{-1} A_0 \exp\left(\xi \frac{t}{\tau_K}\right), \quad (6)$$

i.e., the volume of the consumed fluid grows exponentially with time, contrary to the front volume, i.e., a volume bounded by the leading and trailing edges of the front,

$$V_T(t) \propto A_0 \Delta_F(t) \propto A_0 u' t, \quad (7)$$

whose growth rate is controlled by large-scale turbulent eddies. The latter volume grows linearly with time at $0 < t \ll \tau_T$ or even slower at $\tau_T \leq t$ (Taylor, 1935). At $\xi t/\tau_K = O(1)$, the volume $V_F(t)$ is much less than the volume $V_T(t)$ if $u_0 \ll u_K \ll u'$. Nevertheless, the exponentially growing volume $V_F(t)$ will become larger than the linearly growing $V_T(t)$ at certain instant. This instant could be estimated invoking the following criterion

$$V_F(t^*) = \alpha V_T(t^*), \quad (8)$$

where a coefficient $\alpha < 1$ is introduced in order for some amount of fluid within the volume to remain unburned at $t = t^*$.

It is worth noting that the criterion given by Eq. (8) may be rewritten in the following way

$$\frac{V_F(t^*)}{A_F(t^*)} = \bar{l}(t^*) = \alpha \bar{d}(t^*) = \alpha \frac{V_T(t^*)}{A_F(t^*)}, \qquad (9)$$

where $\bar{l}(t)$ is the mean distance between neighbor elements of the material and self-propagating surfaces and $\bar{d}(t)$ is the mean distance between opposed elements of either the material or the self-propagating surface. The mean distance $\bar{l}(t) = V_F(t)/A_F(t)$ given by Eqs. (3) and (6) is simply equal to

$$\bar{l}(t) \approx \xi^{-1} u_0 \tau_K = \xi^{-1}(u_0/u_K)\eta_K \qquad (10)$$

and is much less than the Kolmogorov length scale if $u_0 \ll u_K \ll u'$. This estimate agrees with the DNS data by Yeung et al. (1990), thus, indicating consistency of the present analysis.

The distance $l$ whose values could randomly be distributed within a wide range of length scales plays the key role in the evolution of the self-propagating surface. Indeed, if the instantaneous distances $d$ and $l$ are locally close to one another, mutual collisions and annihilations of elements of the self-propagating surface appear to be highly probable, with such events drastically reducing the surface area. After annihilation of significant amount of the surface elements, the remaining self-propagating surface, which may be disconnected in a general case, could be close to a small part of the initial material surface. However, the largest part of the remaining self-propagating surface appears to be far from the initial material surface, because the mean $x$-coordinate of the former surface changes with time due to consumption of the fluid, whereas the material surface randomly moves around its constant mean position in the coordinate framework attached to the flow.

As discussed in the next section, annihilation of the front elements takes some time, from $t^*$ to $t^* + \Delta t$. Consequently, Eqs. (3)-(7) do not hold at $t > t^*$. Because a statistical analysis of the front evolution at $t > t^*$ is a complicated task, let us restrict ourselves to estimating the instant $t^*$ and mean characteristics of the front at that instant.

Substitution of Eqs. (6) and (7) into Eq. (8) or substitution of Eqs. (1), (3), (6), and (7) into Eq. (9) yields

$$\alpha (u'/u_0)(\xi t^*/\tau_K) \approx \exp(\xi t^*/\tau_K). \tag{11}$$

Taking logarithm of Eq. (11), we arrive at

$$\xi t^*/\tau_K \approx \ln(\alpha u'/u_0) + \ln(\xi t^*/\tau_K). \tag{12}$$

Under the considered conditions of $u_0 \ll u_K \ll u'$, term $\ln(\alpha u'/u_0) \gg 1$. Therefore, $\xi t/\tau_K \gg 1$, the last term on the right-hand side of Eq. (12) may be neglected when compared to the term on the left-hand side, and approximate solution to the non-linear Eq. (11) reads

$$t^* \approx \xi^{-1} \tau_K \ln(\alpha u'/u_0) \approx \xi^{-1} \tau_T \operatorname{Re}_L^{-1/2} \ln(\alpha u'/u_0). \tag{13}$$

In order for the time $t^*$ given by Eq. (13) to satisfy the constraint of $t \ll \tau_T$, required for the validity of Eqs. (2) and (4), the following estimate should hold

$$\ln(u'/u_0) \ll \operatorname{Re}_L^{1/2}. \tag{14}$$

At instant $t^*$, the front area given by Eqs. (3) and (11) is equal to

$$A_F(t^*)/A_0 = \xi \alpha (u'/u_0)(t^*/\tau_K) \approx \alpha(u'/u_0)\ln(\alpha u'/u_0), \tag{15}$$

the turbulent consumption velocity is equal to

$$u_T(t^*) = u_0(A_F(t^*)/A_0) = \xi \alpha (t^*/\tau_K)u' \approx \alpha u' \ln(\alpha u'/u_0), \tag{16}$$

and the volume of the consumed fluid, given by Eqs. (7) and (8), is equal to

$$V_F(t^*) = \alpha A_0 u' t^*. \tag{17}$$

Finally, the mean consumption velocity averaged over the time interval of $0 < t < t^*$ is equal to

$$\bar{u}_T(t^*) = \frac{V_F(t^*)}{t^*} = \alpha u' \propto u'. \tag{18}$$

and the mean displacement of the front is equal to

$$\overline{X}_F(t^*) = \frac{V_F(t^*)}{A_0} = \alpha u' t^* \approx \alpha \xi^{-1} u' \tau_K \ln(\alpha u'/u_0) \propto \alpha \xi^{-1} u' \tau_T \operatorname{Re}_L^{-1/2} \ln(\alpha u'/u_0). \qquad (19)$$

Under the considered conditions of $u_0 \ll u_K \ll u'$, Eq. (19) shows that $\overline{X}_F(t^*)$ is much larger than the Kolmogorov length scale $\eta_K \ll u'\tau_K \ll \overline{X}_F(t^*)$. Moreover, if Eq. (14) holds, then, $\overline{X}_F(t^*) \ll L$.

Thus, the above analysis of a simple model problem, which is relevant to the influence of the Kolmogorov turbulence on a passive self-propagating front, shows that the mean consumption velocity $\overline{u}_T$ can be proportional to the rms turbulent velocity $u'$, which characterizes large-scale eddies, even if an increase in a ratio of $u_T/u_0$ is mainly controlled by creation of the front area by the smallest Kolmogorov eddies whose scales depend solely on the mean dissipation rate $\overline{\varepsilon}$ and viscosity $\nu$. The straightforward dependence of $\overline{u}_T$ on the Kolmogorov velocity, length, and time scales vanishes due to the following physical mechanism. In intense turbulence characterized by a high Reynolds number, turbulent stretching created by small-scale eddies increases the front area exponentially with time, whereas the volume that the front resides in grows significantly slower. The volume growth rate is controlled by large-scale turbulent eddies and is proportional to the rms dispersion $\Delta_M(t)$, which in its turn is proportional to $t$ or $\sqrt{t}$ at $t \ll \tau_T$ or $\tau_T \ll t$, respectively.

Therefore, the volume is rapidly filled by the front and the mean distance between opposed front elements is rapidly reduced with time, thus, making consumption of the front area due to collisions of these elements highly probable.

Independence of the mean consumption velocity on the Kolmogorov scales does not mean that the Kolmogorov eddies are unimportant. On the contrary, it is the Kolmogorov eddies that create front surface within the framework of the above analysis. Nevertheless, the outcome, i.e., the mean $\bar{u}_T$, is independent of the Kolmogorov scales. This apparent paradox is basically similar to well-known independence of the mean dissipation rate on viscosity in the Kolmogorov turbulence at $\text{Re}_L \to \infty$ or independence of the mean rate of entrainment of ambient irrotational fluid into turbulent fluid on viscosity in shear flows (Townsend, 1976). While both the dissipation and entrainment occur due to viscosity, the mean rates of the two processes are controlled by large-scale velocity fluctuations at $\text{Re}_L \to \infty$, whereas small-scale phenomena adjust themselves to these mean rates. As noted by Tsinober (2009), "*small scales do the 'work', but the amount of work is fixed by the large scales in such a way that the outcome is independent of viscosity*".

**Discussion**

First, it is worth remembering that an analysis of the cyclic behavior of integral characteristics of premixed turbulent combustion such as consumption velocity and flame brush thickness was first performed by Klimov (1975,1983) following an earlier idea by Shelkin that an increase in the instantaneous flame surface area by turbulent eddies should be followed by annihilation of colliding flame elements when the flame surface became highly wrinkled. However, contrary to the present study, Klimov (1983) considered large-scale wrinkles of the flame by eddies of a single velocity scale $u'$ and a single length scale $L$.

Second, while, in order to obtain analytical results, the contents of the previous section was restricted to an early stage of front propagation, the highlighted physical mechanism appears to be of importance during all subsequent stages of the front evolution. Indeed, due to the exponential growth of the front area and a significantly slower growth of the front volume, a period of a rapid growth of the front area should be ended by an annihilation phase when the volume is tightly filled by the front. Such transient effects appear to play a substantial role even during the fully-developed stage of the front propagation. For instance, fully-developed consumption velocity evaluated in a DNS study (Yu et al., 2015) of self-propagation of an infinitely thin passive interface in constant-density turbulence exhibits substantial oscillations with time.

Thus, the following speculations about further evolution of the front appear to be justified. After annihilation of the front elements during a time interval $(t^*, t^* + \Delta t)$, where $\Delta t$ has not yet been estimated, the survived front surface is close to a small part of the initial material surface and the area $A_F(t^* + \Delta t)$ of the former surface is significantly less than $A_F(t^*)$. Then, a new cycle begins. The area of the survived front grows exponentially, see Eq. (3) where $A_0$ is substituted with $A_F(t^* + \Delta t)$, while the dispersion of the survived front relative to its mean coordinate $\overline{X}_F(t^*)$ increases significantly slower. Subsequently, the cycle is completed due to annihilation of colliding front elements and the next cycle begins. Within the framework of the above analysis, the oscillation period could roughly be estimated using Eq. (13), but another physical mechanism could make the period substantially longer.

The point is that the above analysis is based on a hypothesis (supported by DNS data by Yeung et al. (1990)) that self-propagating and material surfaces are close to one another in intense turbulence ($u_0 \ll u_k$) at $t < t^*$. However, even if the front speed $u_0$ is much less than the Kolmogorov velocity $u_K$, the front can move at a speed much larger than $u_K$ due to formation of cusps with a very acute angle. For instance, appearance of cusps in premixed turbulent flames was addressed in a recent DNS study by Poludnenko and Oran (2011) and cusp-like structures of various (and even large) length scales are clearly visible in DNS images of material surfaces, reported by Goto and Kido (2007). Due to rapid motion of a cusp with a very acute angle, elements of initially coinciding self-propagating and material surfaces can diverge. Accordingly, Eq. (3) may not hold in the vicinity of cusps and the rate of growth of the front area may be reduced. Moreover, while the surface area is consumed near a cusp, this process may be sufficiently long (when compared to $t^*$) provided that the cusp is sufficiently elongated. Accordingly, appearance of cusps may increase $\Delta t$ and the period of the cycles discussed earlier. The issue definitely needs further study.

Finally, it is worth stressing that the transient effects highlighted in the present letter are fundamentally different from oscillations of the turbulent flame speed and mean flame brush thickness due to the growth and disappearance of unburned mixture fingers (Lipatnikov et al., 2015; Poludnenko, 2015; Sabelnikov and Lipatnikov, 2017). The point is that the finger growth is controlled by pressure gradient generated due to combustion-induced thermal expansion (Lipatnikov et al., 2015), but such thermal expansion effects vanish in the constant-density case analyzed above.

**Conclusions**

When small-scale turbulent eddies stretch a slowly ($u_0 \ll u'$) propagating front and increase its area, such an increase in the area cannot be continuous long. Due to the exponential growth of the area, the front packing in the front volume is limited by annihilation of the front elements in mutual collisions. Accordingly, a stage characterized by rapidly growing front area and consumption velocity should be followed by a stage during that the area partly disappears and the velocity drops. Due to this physical mechanism, transient effects (oscillations) could play a substantial role even during fully-developed stage of the front propagation. Moreover, due to this physical mechanism and the transient effects caused by it, the mean turbulent consumption velocity $\bar{u}_T$ may adjust itself to the rate of turbulent entrainment, i.e., to the rms turbulent velocity $u'$, which characterizes large-scale eddies. The smallest eddies of the Kolmogorov scales do not affect the mean area of the front and the turbulent consumption velocity, respectively, in spite of the fact that an increase in the front area and, hence, an increase in a ratio of $u_T(t)/u_0$ are mainly controlled by such eddies. In some sense, the Kolmogorov eddies behave like Cheshire cat from *Alice in Wonderland*.

**Funding**

This work was supported by ONERA, by the Grant of the Ministry of Education and Science of the Russian Federation (Contract No. 14.G39.31.0001 of 13.02.2017), by Chalmers Combustion Engine Research Center (CERC), and by Chalmers Transport Area of Advance.

**References**

Aris, R. 1999. *Mathematical Modeling. A Chemical Engineer's Perspective,* Academic Press, New York, NY.

Batchelor, G.K. 1952. The effect of homogeneous turbulence on material lines and surfaces. *Proc. R. Soc. London A,* **213**, 349-366.

Chaudhuri, S., Wu, F., Zhu, D., and Law, C.K. 2012. Flame speed and self-similar propagation of expanding turbulent premixed flames. *Phys. Rev. Lett.*, **108**, 044503.

Chertkov, M., and Yakhot, V. 1998. Propagation of a Huygens front through turbulent medium. *Phys. Rev. Lett.*, **80**, 2837-2840.

Damköhler, G.Z. 1940. Der einfuss der turbulenz auf die flammengeschwindigkeit in gasgemischen. *Electrochem. Angew. Phys. Chem.*, **46**, 601-652.

Gamezo, V.N., Khokhlov, A.M., and Oran, E.S. 2004. Deflagrations and detonations in thermonuclear supernovae. *Phys. Rev. Lett.*, **92**, 211102.

Gamezo, V.N., Khokhlov, A.M., Oran, E.S., Chtchelkanova, A.Y., and Rosenberg, R.O. 2003. Thermonuclear supernovae: simulations of the deflagration stage and their implications. *Science*, **299**, 77-81.

Girimaji, S.S., and Pope, S.B. 1990. Propagating surfaces in isotropic turbulence. *J. Fluid Mech.*, **220**, 247-277.

Goto, S., and Kida, S. 2007. Reynolds-number dependence of line and surface stretching in turbulence: folding effects. *J. Fluid Mech.*, **586**, 59-81.

Kerstein, A.R., and Ashurst, W.T. 1992. Propagation rate of growing interfaces in stirred fluids. *Phys. Rev. Lett.,* **68**, 934-937.

Klimov, A.M. 1975. Flame propagation in intense turbulence. *Dokl. Akad. Nauk SSSR*, **221**, 56-59.

Klimov, A.M. 1983. Premixed turbulent flames - interplay of hydrodynamic and chemical phenomena. *AIAA Prog. Astronaut. Aeronaut.*, **88**, 133-146.

Kolmogorov, A.N. 1941. The local structure of turbulence in incompressible viscous fluid for very large Reynolds number. *Dokl. Akad. Nauk SSSR*, **30**, 299-303 [English translation *Proc. R. Soc. London A*, **434**, 9-13, 1991].

Lipatnikov, A.N. 2012. *Fundamentals of Premixed Turbulent Combustion*, CRC Press, Boca Raton, FL.

Lipatnikov, A.N., and Chomiak, J. 2002. Turbulent flame speed and thickness: phenomenology, evaluation, and application in multi-dimensional simulations. *Prog. Energy Combust. Sci.*, **28**, 1-73.

Lipatnikov, A.N., Chomiak, J., Sabelnikov, V.A., Nishiki, S., and Hasegawa, T. 2015. Unburned mixture fingers in premixed turbulent flames. *Proc. Combust. Inst.*, **35**, 1401-1408.

Mayo, J.R., and Kerstein, A.R. 2008. Fronts in randomly advected and heterogeneous media and nonuniversality of Burgers turbulence: Theory and numerics. *Phys. Rev. E,* **78**, 056307.

Niemeyer, J.C., and Kerstein, A.R. 1997. Numerical investigation of scaling properties of turbulent premixed flames. *Combust. Sci. Technol*., **128**, 343-358.

Peters, N. 1986. Laminar flamelet concepts in turbulent combustion. *Proc. Combust. Inst*., **21**, 1231-1249.

Poinsot, T., and Veynante, D. 2005. *Theoretical and Numerical Combustion*, 2nd ed., Edwards, Philadelphia, PA.

Poludnenko, A.Y. 2015. Pulsating instability and self-acceleration of fast turbulent flames. *Phys. Fluids*, **27**, 014106.

Poludnenko, A.Y., and Oran, E.S. 2011. The interaction of high-speed turbulence with flames: Turbulent flame speed. *Combust. Flame,* **158**, 301-326.

Poludnenko, A.Y., Gardiner, T.A., and Oran, E.S. 2011. Spontaneous transition of turbulent flames to detonations in unconfined media. *Phys. Rev. Lett*., **107**, 054501.

Sabelnikov, V.A., and Lipatnikov, A.N. 2017. Recent advances in understanding of thermal expansion effects in premixed turbulent flames. *Annu. Rev. Fluid Mech*., **49**, 91-117.

Shelkin, K.I. 1943. On combustion in a turbulent flow. *Zhournal Tekhnicheskoi Fiziki*, **13**, 520-530 [English translation NACA TM 1110, 1947].

Shy, S.S., Ronney, P.D., Buckley, S.G., and Yakhot, V. 1992. Experimental simulation of premixed turbulent combustion using aqueous autocatalytic reactions. *Proc. Combust. Inst.,* **24**, 543-551.

Sreenivasan, K.R., Ramshankar, R., and Meneveau, C. 1989. Mixing, entrainment and fractal dimensions of surfaces in turbulent flows. *Proc. R. Soc. London A*, **421**, 79-108.

Taylor, G.I. 1935. Statistical theory of turbulence. IV. Diffusion in a turbulent air stream. *Proc. R. Soc. London A*, **151**, 421-478.

Townsend, A.A. 1976. *The Structure of Turbulent Shear Flow*, 2nd ed., Cambridge University Press, Cambridge, UK.

Tsinober, A. 2009. *An Informal Conceptual Introduction to Turbulence,* Springer, Heidelberg, Germany.

Yeung, P.K., Girimaji, S.S., and Pope, S.B. 1990. Straining and scalar dissipation of material surfaces in turbulence: Implications for flamelets. *Combust. Flame***, 79**, 340-365.

Yu, R., and Lipatnikov, A.N. 2017a. Direct numerical simulation study of statistically stationary propagation of a reaction wave in homogeneous turbulence. *Phys. Rev. E*, **95**, 063101.

Yu, R., and Lipatnikov, A.N. 2017b. DNS study of dependence of bulk consumption velocity in a constant-density reacting flow on turbulence and mixture characteristics. *Phys. Fluids*, **29**, 065116.

Yu, R., Bai, X.-S., and Lipatnikov, A.N. 2015. A direct numerical simulation study of interface propagation in homogeneous turbulence. *J. Fluid Mech*., **772**, 127-164 (2015).